\documentclass{emulateapj}
\usepackage{bm}
\usepackage{color}

\newcommand{\beq}{\begin{equation}}
\newcommand{\beqa}{\begin{eqnarray}}
\newcommand{\eeq}{\end{equation}}
\newcommand{\eeqa}{\end{eqnarray}}

\newcommand{\bfx}{\mathbf{x}}

\shorttitle{Revising the Halofit Model}
\shortauthors{Takahashi et al.}

\begin{document}

\title{Revising the Halofit Model for the Nonlinear Matter Power Spectrum}

\author{Ryuichi Takahashi\altaffilmark{1},
Masanori Sato\altaffilmark{2},
Takahiro Nishimichi\altaffilmark{3},
Atsushi Taruya\altaffilmark{3,4}, and
Masamune Oguri\altaffilmark{3}
}

\affil{\altaffilmark{1} Faculty of Science and Technology, Hirosaki
 University, 3 bunkyo-cho, Hirosaki, Aomori, 036-8561, Japan}
\affil{\altaffilmark{2} Department of Physics,
 Nagoya University, Chikusa, Nagoya 464-8602, Japan}
\affil{\altaffilmark{3} Kavli Institute for the Physics and Mathematics of the
 Universe, Todai Institutes for Advanced Study, the University of Tokyo,
 Kashiwa, Japan 277-8583 (Kavli IPMU, WPI)}
\affil{\altaffilmark{4} Research Center for the Early Universe, 
 The University of Tokyo, Tokyo 133-0033, Japan}

\begin{abstract}
Based on a suite of state-of-the-art high-resolution $N$-body
simulations, we revisit the so-called halofit model (Smith et
al. 2003) as an accurate fitting formula for the nonlinear matter
power spectrum. While the halofit model has been frequently used as a
standard cosmological tool to predict the nonlinear matter power
spectrum in a universe dominated by cold dark matter, its precision
has been limited by the low-resolution of $N$-body simulations used to
determine the fitting parameters, suggesting the necessity of improved
fitting formula at small scales for future cosmological studies. We run
high-resolution $N$-body simulations for 16 cosmological models around
the Wilkinson Microwave Anisotropy Probe (WMAP) best-fit cosmological
parameters (1, 3, 5, and 7 year results), including dark energy models
with a constant equation of state. The simulation results are used to
re-calibrate the fitting parameters of the halofit model so as to
reproduce small-scale power spectra of the $N$-body simulations, while
keeping the precision at large scales. The revised fitting formula
provides an accurate prediction of the nonlinear matter power spectrum
in a wide range of wavenumber ($k \leq 30h$\,Mpc$^{-1}$) at
 redshifts $0 \leq z \leq 10$, with $5\%$~precision for
 $k\leq1\,h$\,Mpc$^{-1}$ at $0 \leq z \leq 10$ and $10\%$ for 
 $1 \leq k\leq 10\, h$\,Mpc$^{-1} $ at $0 \leq z \leq 3$. 
We discuss the
impact of the improved halofit model on weak lensing power spectra and
correlation functions, and show that the improved model better
reproduces ray-tracing simulation results. 
\end{abstract}


\section{Introduction}

The large-scale structure of the Universe has evolved under the
influence of cosmic expansion and gravity, and its statistical nature
contains valuable cosmological information.  
Among others, the power spectrum $P(k)$ is one of the most fundamental
statistical quantities characterizing the large-scale 
structure. It has widely been used for cosmological studies, both in
predicting various observable quantities and in extracting
cosmological information from the observations~\citep[e.g.,][]{pee93,dod03}.   
Given growing interests in high precision cosmological observations, 
of particular importance is an accurate theoretical template of the 
power spectrum, taking account of the nonlinear gravitational
evolution. 

Weak lensing induced by the large-scale structure between observed
galaxies and the observer provides a unique opportunity to directly
probe matter inhomogeneities in the Universe. This cosmic shear signal
has been measured with a high signal-to-noise ratio by current large
surveys including Canada-France-Hawaii Telescope Legacy
Survey~\citep[CFHTLS;][]{fu08}, Sloan Digital Sky
Survey~\citep[SDSS;][]{lin11,huff11}, and Cosmic Evolution
Survey~\citep[COSMOS;][]{massey07,sch10}. These surveys provided
useful constraints on the cosmological parameters such as the matter
density parameter $\Omega_{\rm m}$ and the amplitude of density
fluctuation $\sigma_8$. Future surveys such as Subaru Hyper
Suprime-Cam~\citep[HSC;][]{miyazaki06}, 
Dark Energy Survey~\citep[DES;][]{des05}, and 
Large Synoptic Survey Telescope~\citep[LSST;][]{lsst09} aim at measuring 
the cosmic shear signal with unprecedented precisions. While weak
lensing probes matter fluctuations projected along the line-of-sight, 
one can extract the redshift evolution of the fluctuations, and hence
accurate information on dark energy, using a technique called lensing
tomography~\citep[e.g.,][]{Hu99,TJ04} or a cross-correlation with
intervening objects \citep[e.g.,][]{oguri11}. However, accurate and
unbiased cosmological constraints from these lensing measurements can
be obtained only if we have appropriate likelihood function with given
marginal distributions~\citep{sit10,sit11}
and an accurate model of the power spectrum
$P(k)$. For instance, \citet{ht05} argued that we typically need a few
percent accuracy of $P(k)$ at the wavenumber $k<10h$\,Mpc$^{-1}$ in
order for the uncertainty of $P(k)$ not to degrade cosmological
constraints in DES and LSST ~\citep[see also][in which a similar
  conclusion is obtained]{eif11,hzm12}. 

In the linear and quasi-linear regime of density fluctuations, the
power spectrum can be computed for any given initial conditions and
cosmological parameters using perturbation theory~\citep[e.g.,][for
  a review]{ber02}. In the nonlinear regime, however, one has to
resort to cosmological $N$-body simulations to study the nonlinear
gravitational evolution. $N$-body simulation results are then used
to develop phenomenological halo models or fitting formulae of 
nonlinear gravitational clustering. For instance, \citet{pd96}
provided a fitting formula of $P(k)$ based on a scaling ansatz
presented in \citet{hklm91}. \citet[][hereafter S03]{sm03} proposed
a new model of $P(k)$, the so-called halofit model, which is based 
on a halo model of structure formation \citep[e.g.,][]{mf00,sel00,CS02}.
In this halo model, all the matter content in the Universe is assumed
to be bound in dark matter halos. Then the power spectrum is
decomposed into two terms, the so-called one- and two-halo terms. The
one-halo term describes matter correlations within the same dark
matter halo, and is determined by the density profile of each halo. On
the other hand, the two-halo term arises from the correlation between
two distinct halos. The one-halo term  dominates at small scales,
whereas the two-halo term dominate at large scales. The halofit model
chose the functional form of $P(k)$ based on the halo model, but the
model parameters were calibrated from $N$-body simulation results.

The halofit model by S03 is widely used to calculate the nonlinear
matter power spectrum, yet it has been reported that the model fails
to reproduce recent high-resolution $N$-body simulation results at
small scales~\citep[e.g.,][]{spr05,hil09,sato09,boy09,taka11,kie11,vn11a,vn11b,har12,cas12, it12}. 
For instance, \citet{wv04} first pointed out that the halofit
predicts a smaller power than their numerical results at small scales.
\citet{heit10} ran a suite of high-resolution simulations,
called ``Coyote Universe'', and showed that $P(k)$ predicted by the
halofit is $\sim 5\%$ smaller than their numerical results at 
$k \sim 1h$\,Mpc$^{-1}$. The one reason of the difference comes from the
fact that the $N$-body simulations used in S03 have lower spatial
resolution than latest ones.
The another reason is that the halofit model in S03 is the fitting 
 function for the Cold Dark Matter (CDM) model without
 baryons\footnote{The presence of a significant fraction of baryon
 suppress the linear power spectrum at small scales.
 The fitting function in S03 is evaluated from the input
 linear power spectrum. Hence, the fitting function 
 is slightly biased for the cosmological models with baryons.}.
An outcome of the Coyote Universe
simulations is a publicly available code ``cosmic emulator'' to
calculate the nonlinear matter power spectrum by interpolating the
simulations results for 38 different cosmological models \citep{law10}.
However, their emulator is restricted to a narrow range in $k<3h{\rm
  Mpc}^{-1}$ and at low redshift $0 \le z \le 1$. 
Also, the Hubble parameter is automatically specified in the code
using the cosmic microwave background (CMB) anisotropy constraint on
the distance to the last scattering surface. 

In this paper, we revisit the halofit model based on state-of-the-art
high-resolution $N$-body simulations in 16 cosmological models around
the Wilkinson Microwave Anisotropy Probe (WMAP) best-fit cosmological
parameters. We allow the dark energy equation of state $w$ to deviate
from $-1$, assuming that $w$ does not evolve with redshift.
The halofit model has been tested for dark energy models
 ($w \neq -1$) using N-body simulations
 ~\citep[e.g.,][]{mtc06,ma07,cas09,fll09,ali10,cas11b}.
While the
original halofit model in S03 contains $30$ parameters, we increase
the number of parameters to $35$ in order to achieve a better fit to
the simulations. The new formula we present, which is summarized in
Appendix, is widely applicable in the wavenumber range of 
$k<30h$\,Mpc$^{-1}$ and the redshift range of $0 \leq z \leq 10$.
Simply replacing the parameters in the original halofit model with 
new ones in the Appendix, an accuracy of fitting function is improved 
especially at small scales.  

The present paper is organized as follows. In Section~\ref{sec:nbody}, 
we begin by describing the $N$-body simulations and cosmological
models used for the power spectrum analysis. Combining the 
$N$-body results with different box sizes, we discuss in detail 
the convergence of power spectrum measurement over the wide 
range of wave number. In Section~\ref{sec:halo}, we re-calibrate the
halofit model, and the revised version of the halofit model, whose
explicit formula is given in Appendix, is compared with our
$N$-body simulations. As an important implication of revised halofit
model, in Section~\ref{sec:wl}, we compute weak lensing power
spectra, and compare them with direct ray-tracing simulation results,
particularly focusing on the small-scale behavior. Finally,
Section~\ref{sec:conc} is devoted to conclusion and discussion. 

\section{N-body Simulations}\label{sec:nbody}

\begin{deluxetable}{lllllll}
\startdata
  \hline
  & $\Omega_{\rm b}$  & $\Omega_{\rm m}$ & $h$ & $\sigma_8$ & $n_{\rm s}$
 & $-w$ \\ \hline 
  WMAP1 & $0.044$ & $0.29$  & $0.72$ & $0.9$ & $0.99$ & $1$ \\
  WMAP3 & $0.041$ & $0.238$ & $0.732$ & $0.76$ & $0.958$ & $1$ \\
  WMAP5 & $0.046$ & $0.279$ & $0.701$ & $0.817$ & $0.96$ & $1$ \\
  WMAP7 & $0.046$ & $0.272$  & $0.7$ & $0.81$ & $0.97$ & $1$  \\
  WMAP7a & $0.046$ & $0.272$  & $0.7$ & $0.81$ & $0.97$ & $0.8$  \\
  WMAP7b & $0.046$ & $0.272$  & $0.7$ & $0.81$ & $0.97$ & $1.2$
\enddata
\tablecomments{
 Best-fit cosmological parameters in a series of WMAP papers. Here we
 show the baryon density $\Omega_{\rm b}$, the matter density
 $\Omega_{\rm m}$, the Hubble constant $h$, the amplitude of power
 spectrum at $8h^{-1}$\,Mpc $\sigma_8$, the spectral index $n_{\rm s}$,
 and the equation of state of dark energy $w$. 
 We assume a flat curvature ($\Omega_{\rm w}=1-\Omega_{\rm m}$).
}
\label{table1}
\end{deluxetable}

\begin{deluxetable}{lllllll}
\startdata
  \hline
  & $\Omega_{\rm b}$  & $\Omega_{\rm m}$ & $h$ & $\sigma_8$ & $n_{\rm s}$
  & $-w$ \\ \hline 
  m00 & $0.0432$ & $0.25$  & $0.72$ & $0.8$ & $0.97$ & $1$ \\
  m01 & $0.0647$ & $0.4307$ & $0.5977$ & $0.8161$ & $0.9468$ & $0.816$ \\
  m02 & $0.0637$ & $0.4095$ & $0.5907$ & $0.8548$ & $0.8952$ & $0.758$ \\
  m03 & $0.0514$ & $0.2895$  & $0.6763$ & $0.8484$ & $0.9984$ & $0.874$  \\
  m04 & $0.0437$ & $0.2660$  & $0.7204$ & $0.7$ & $0.9339$ & $1.087$  \\
  m05 & $0.0367$ & $0.2309$  & $0.7669$ & $0.8226$ & $0.9726$ & $1.242$ \\
  m06 & $0.0462$ & $0.3059$  & $0.7040$ & $0.6705$ & $0.9145$ & $1.223$ \\
  m07 & $0.0582$ & $0.3310$  & $0.6189$ & $0.7474$ & $0.921$ & $0.7$  \\
  m08 & $0.0428$ & $0.2780$  & $0.7218$ & $0.8090$ & $0.9855$ & $1.203$  \\
  m09 & $0.0623$ & $0.3707$  & $0.6127$ & $0.6692$ & $0.979$ & $0.739$ 
\enddata
\tablecomments{
Cosmological parameters of Coyote models. 
}
\label{table2}
\end{deluxetable}

\begin{deluxetable*}{lccccccl}
\tabletypesize{\scriptsize}
\startdata
  \hline
  & $L (h^{-1}{\rm Mpc})$ & $N_p^3$ & $N_{\rm r}$ &
 $k_{\rm Nyq} (h \,{\rm Mpc}^{-1})$ & $r_s (h^{-1} {\rm kpc})$
 & $z_{\rm init}$ &
 \hspace{1.8cm} $z_{\rm out}$   \\ \hline 
  WMAP models & $2000$ & $1024^3$ & $3$ & $1.6$ & $97.6$ & $99$ &
 $0,~0.35,~0.7,~1,~1.5,~2.2,~3,~5,~7,~10$ \\
              & $800$ & $1024^3$ & $3$ & $4.0$ & $39.0$ & $99$ &
 $0,~0.35,~0.7,~1,~1.5,~2.2,~3,~5,~7,~10$ \\
              & $320$ & $1024^3$ & $6$ & $10.$ & $15.6$ & $99$ &
 $0,~0.35,~0.7,~1,~1.5,~2.2,~3,~5,~7,~10$ \\
  \hline
  Coyote models & $1000$ & $1024^3$ & $1$ & $3.2$ & $48.8$ & $99$ &
 $0,~0.35,~0.7,~1,~1.5,~2.2,~3,~5,~7,~10$ \\
    & $320$ & $1024^3$ & $1$ & $10.$ & $15.6$ & $99$ &
 $0,~0.35,~0.7,~1,~1.5,~2.2,~3,~5,~7,~10$ 
\enddata
\tablecomments{
 Model parameters of our numerical simulations for the WMAP models
 (upper rows) and the Coyote models (lower rows):
 the box size $L$, the number of particles $N_p^3$,
 the number of realizations $N_{\rm r}$,
 the Nyquist frequency $k_{\rm Nyq}=(2 \pi/L) (N_p/2)$, the softening
 length $r_s$, the initial
 redshift $z_{\rm init}$ and the redshifts of the simulation outputs
 $z_{\rm out}$.
}
\label{table3}
\end{deluxetable*}

\subsection{Power Spectrum}\label{sec:ps}

In this section, we describe our cosmological $N$-body simulations
used in this paper. We follow the nonlinear gravitational evolution
of $1024^3$ collisionless particles in a cubic box of side $L$.
We use the public cosmological $N$-body simulation code Gadget2 which
is a tree-PM code~\citep{syw01,s05}.  We use $2048^3$ PM grid to
follow the gravitational evolution at small scales accurately.
We generate the initial conditions based on the second-order
Lagrangian perturbation theory~\citep[2LPT;][]{cps06,n09} with
the initial linear power spectrum calculated by the Code for
Anisotropies in the Microwave Background~\citep[CAMB;][]{lcl00}.
The initial redshift is set to $z_{\rm in}=99$. We store simulation
results (particle positions) at various redshifts from $z=0$ to
$10$. The softening length is set to $5\%$ of the mean particle
separation. To calculate the power spectrum, we assign the particles
on $N_g^3=1280^3$ grid points using the cloud-in-cells (CIC)
method~\citep{he81} to obtain the density field. After performing the
Fourier transform, we correct the window function of CIC by dividing
each mode by the Fourier transform of the window kernel as
$\tilde{\delta}_{\bf k} \rightarrow \prod_{i=x,y,z} \left[{\rm sinc}
  (L k_i/2 N_{\rm g})\right]^{-2} \times \tilde{\delta}_{\bf k}$,
where $\tilde{\delta}_{\bf k}$ is the density fluctuation in Fourier
space and ${\rm sinc}(x)=\sin(x)/x$~\citep[e.g.,][]{taka09,sm11}. 
In addition, to evaluate the power spectrum at small scales accurately,
we fold the particle positions into a smaller box by replacing  
$\bfx \rightarrow \bfx \% (L/2^n)$ where the operation $a \% b$ stands
for the remainder of the division of $a$ by
$b$~\citep[e.g.,][]{jenk98, sm03, vn11a}.
This procedure leads to effectively $2^n$ times higher resolution. 
Here we adopt $n=0$, $2$, and $4$. We use the density fluctuation
$\tilde{\delta}_{\bf k}$ up to half the Nyquist frequency determined
by the box size $L/2^n$ with the grid number $N_g$, i.e., 
$1/2 \times k_{\rm Nyq}$ $=(\pi/(L/2^n))(N_g/2)$, with $n=0$, $2$, and
$4$. This condition corresponds to $k<6.3$, $25$, and $100
h$\,Mpc$^{-1}$ with $n=0$, $2$, and $4$, respectively, for the box
size of $L=320 h^{-1}$Mpc with $N_g=1280$. Finally, we compute the
power spectrum 
\beq
  P(k) =  \frac{1}{N_k} \sum_{\bf k}\left| \tilde{\delta}_{\bf k} \right|^2,
\eeq
where the summation over Fourier modes is done for the modes falling
into the bin [$k-\Delta k/2$,~$k+\Delta k/2$], and $N_k$ denotes the
number of available Fourier modes in the bin. We do not subtract the
shot noise in the measured power spectrum. Instead, we do not use
$P(k)$ at small scales where the shot noise dominates (see
Section~\ref{sec:halo}).  

\subsection{Cosmological Models}

In this paper, we use simulation results for 16 cosmological models.
Six are taken from the results of WMAP papers and 10 are from the
cosmological models adopted by the Coyote Universe. For all of the
models, we assume a flat curvature ($\Omega_{\rm w} =1-\Omega_{\rm
  m}$, where $\Omega_{\rm w}$ is the dark energy density). 
The first four WMAP models, which are shown in Table~\ref{table1}, are
the best-fit $\Lambda$CDM models of WMAP 1, 3, 5, 7~year
results~\citep{sper03,sper07,k09,k11}. The other two models, WMAP7a
and WMAP7b, are the same as the WMAP7 model except that we slightly
change the equation of state parameter of dark energy ($w=-0.8$ and
$-1.2$). 

In addition, we also examine 10 models among 38 cosmological 
models presented in the Coyote Universe, as shown in
Table~\ref{table2}. These models, tagged as m00 to m09, were used 
in a series of papers of the Coyote Universe
project~\citep{heit09,heit10,law10}, in which 38 cosmological models
in a parameter range of $0.120 < \Omega_{\rm m} h^2 < 0.155$, 
$0.0215 < \Omega_{\rm b} h^2 < 0.0235$,  
$0.85 < n_{\rm s} < 1.05$, $-1.30 < w < -0.70$, and $0.61 < \sigma_8 < 0.9$
are used to make a fitting function of the power spectrum.
The models used in our paper correspond to their first 10 cosmological
models. Since the cosmological parameters in Coyote models are
different from the WMAP models typically by $20-30 \%$, we use our
simulation results for these Coyote cosmological models to check the
dependence of our fitting function on cosmological parameters.

Table~\ref{table3} summarizes our simulation setting, including the
box size, the number of particles, the number of realizations, and
the softening length, for the WMAP and Coyote models. In the WMAP models,
we adopt the simulation boxes of $L=2000$, $800$, and $320h^{-1}$Mpc
on a side. We prepare $3$ different random realizations for $L=2000$
and $800h^{-1}$Mpc, and $6$ realizations for $320h^{-1}$Mpc to reduce
the sample variance. We combine the power spectrum $P(k)$ obtained
from the different simulation boxes to cover a wide wavenumber
range. The specific procedure for combining the $P(k)$ is discussed in
the next subsection. We use the mean power spectrum of these
realizations. The Nyquist wavenumber of the mean particle separation
in the smallest box ($L=320h^{-1}$Mpc) is $k=10h$\,Mpc$^{-1}$. The
bin width is set linearly, $\Delta k = 0.01h$\,Mpc$^{-1}$, in the
linear regime $k\le 0.3 h$\,Mpc$^{-1}$, and logarithmically, 
$\Delta \log_{10} (k/h$\,Mpc$^{-1}) =0.02$, in the nonlinear regime
$k>0.3 h$\,Mpc$^{-1}$. We analyze ten outputs at redshifts $0 \leq z
\leq 10$, $z=0$, $0.35$, $0.7$, $1$, $1.5$, $2.2$, $3,5$, $7$, and $10$.
We checked that the power spectra of our simulations agree with
the results of higher resolution simulations, in which we set the
finer simulation parameters for the time step, the force calculation,
etc., within $2(6) \%$ for $k<10(30)h$\,Mpc$^{-1}$.\footnote[\#]{The Gadget-2
 parameters we used are ErrTolIntAccuracy $=0.05$, MaxSizeTimeStep $=0.03$,
 MaxRMSDisplacementFac $=0.25$, ErrTolTheta $=0.5$, ErrTolForceAcc $=0.003$,
 and PMGRID $=2048$.}

In the Coyote models, we use the simulation boxes of $1000h^{-1}$Mpc 
 and $320h^{-1}$Mpc on a side. 
We prepare a single realization for each of the 10 models. 
The Nyquist wavenumber is $k=3.2(10)h$\,Mpc$^{-1}$ for
 $L=1000(320)h^{-1}$Mpc and we use a
logarithmic bin of $\Delta \log_{10} (k/h$\,Mpc$^{-1}) =0.1$.
We use ten outputs at the redshifts $0 \leq z \leq 10$, 
$z=0$, $0.35$, $0.7$, $1$, $1.5$, $2.2$, $3$, $5$, $7$, and $10$.
We confirmed that our simulation results agree with the cosmic
emulator results within $3 \%$ for $0.1 < k < 3 h$\,Mpc$^{-1}$ at 
$0 \leq z \leq 1$.

Finally, as a further cross check, we compare our simulation results
with  two high-resolution simulation results in \cite{vn11a} and
\cite{taka11}. In both the simulations, the same codes, Gadget-2 with
2LPT initial condition, were used. \cite{vn11a} employed $2048^3$
particles in different box sizes of $L=4096$, $2048$, $1024$,
and $512h^{-1}$Mpc. They calculated the power spectra at redshifts
$z=0.35$, $1$, and $3$ in the WMAP5 model. The Nyquist wave number in
the smallest box is $13h$\,Mpc$^{-1}$. On the other hand,
\cite{taka11} employed $1024^3$ particles on the box size of
$50h^{-1}$Mpc at redshift $z=0-20$. They prepared independent four
realizations. The cosmological parameters are based on the WMAP 5 year
result, although the values were slightly different from those of the
WMAP5 model listed in Table 1. We use $14$ outputs at redshifts
$z=0-10$, $z=0$, $0.35$, $0.7$, $1$, $1.5$, $2.2$, $3$, $3.4$, $4.1$,
$4.7$, $5.5$, $6.4$, $7.6$, and $9.1$. The Nyquist wave number is 
$64h$\,Mpc$^{-1}$. We use these simulation results in our analysis
for only small scales, $k>10(1)h$\,Mpc$^{-1}$ at $z \leq 3 (>3)$,
in order to check the asymptotic behavior of our fitting formula at
high $k$ limit.  

\subsection{Accuracy of Our $N$-body Simulations}

\begin{figure}
\epsscale{1.1}
\plotone{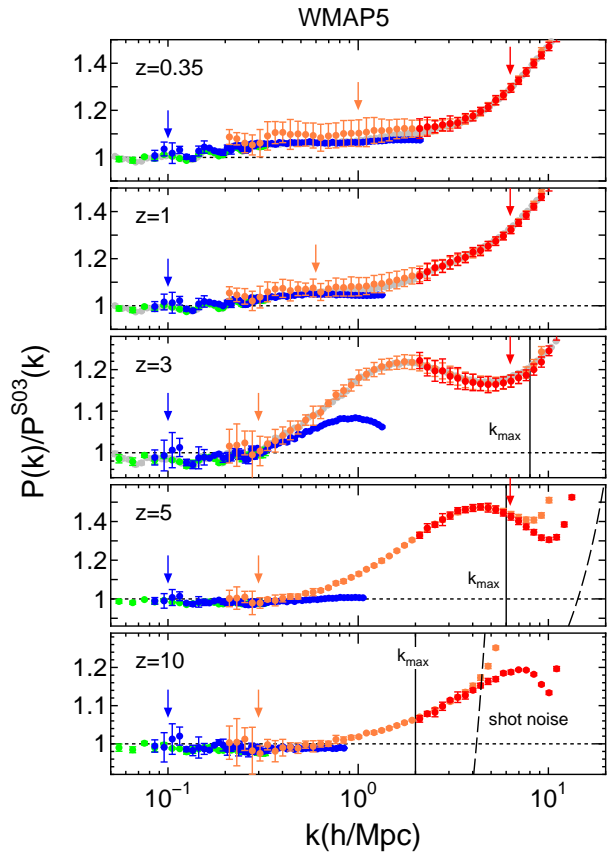}
\caption{
 Power spectra $P(k)$ from the different simulation box sizes
 for the WMAP5 model at redshifts $z=0.35$, $1$, $3$, $5$, and
 $10$. The vertical axis shows power spectra in our simulations
 normalized by the theoretical nonlinear matter power spectrum model
 from the original halofit model in S03. Green, blue, and orange
 symbols are simulation results from the box sizes of $L=2000$, 
 $800$, and $320h^{-1}$Mpc, respectively. Red symbols are the same as
 the oranges, but including the folding method with $n=2$ (see
 Section~\ref{sec:ps}). Gray symbols show simulation results from 
 \cite{vn11a}. Arrows denote wavenumbers where $P(k)$ with the
 different box sizes are connected. Dashed curves denote the shot
 noise. Vertical solid lines indicate the maximum wavenumber $k_{\rm
   max}$ for deriving our fitting formula.}
\label{pk_wmap5}
\vspace*{0.5cm}
\end{figure}

\begin{figure}
\epsscale{1.0}
\plotone{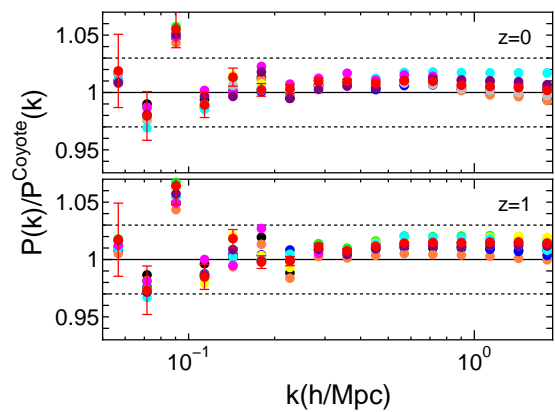}
\caption{
 Power spectra $P(k)$ from our simulations of $L=1000h^{-1}$Mpc
 divided by those of the
 Cosmic emulator for the ten Coyote models at $z=0$ ({\it top}) and $1$
 ({\it bottom}). Red symbols with error bars are for the fiducial
 model m00. Horizontal dotted lines indicate the fractional error of
 $3 \%$. 
}
\label{pk_coyote_m00}
\vspace*{0.5cm}
\end{figure}

\begin{deluxetable}{ccccl}
\startdata
  \hline
  $z$ & $k_{2000/800}$ & $k_{800/320}$ & $k_{\rm min}$ &
 $k_{\rm max}$   \\ \hline 
 $0,0.35$ &  $0.1$ & $1.$ & $0.02$ & $30$     \\
 $0.7,1$ & $0.1$ & $0.6$ & $0.02$ & $20$    \\
 $1.5,2.2$ & $0.1$ & $0.4$ & $0.02$ & $10$    \\
 $3$ & $0.1$ & $0.3$ & $0.02$ & $8$   \\
 $5$ & $0.1$ & $0.3$ & $0.02$ & $6$   \\
 $7$ & $0.1$ & $0.3$ & $0.02$ & $3$   \\
 $10$ & $0.1$ & $0.3$ & $0.02$ & $2$  
\enddata
\tablecomments{
 The summary of wavenumbers (in units of $h$\,Mpc$^{-1}$) where the
 simulation results $P(k)$ with the different box sizes are connected
 at each redshift, $k_{2000/800}$ is for $L=2000$ and $800h^{-1}$Mpc,
 and $k_{800/320}$ is for $L=800$ and $320h^{-1}$Mpc. The minimum and
 maximum wavenumbers used in our analysis are shown by $k_{\rm min}$
 and $k_{\rm max}$, respectively.
}
\label{table4}
\end{deluxetable}

In this subsection, we compare our simulation results with previous
works to check the accuracy and convergence of our $N$-body
simulations. We also describe the procedure for combining the $P(k)$
from the different simulation box sizes. Figure~\ref{pk_wmap5} shows
the power spectrum $P(k)$ of our simulations for the different box
sizes for the WMAP5 model at $z=0.35$, $1$, $3$, $5$, and $10$.  The
vertical axis shows the measured power spectra normalized by the
theoretical model of the nonlinear power spectra from the original
halofit model in S03. Green, blue, and orange symbols are the results
from the different box sizes of $L=2000$, $800$, and $320h^{-1}$Mpc,
respectively. Red symbols are the same simulation results as plotted
in the orange symbols ($L=320h^{-1}$Mpc), but using the folding method
with $n=2$ in computing $P(k)$ (see Section~\ref{sec:ps}). We plot the
mean $P(k)$ with the error bars among the realizations. Gray
symbols at $z=0$, $0.35$, and $1$ are the simulation results of
\cite{vn11a}. Vertical arrows indicate the wavenumbers at which we
connected the simulation results from the different box sizes. For
example, we used the results from $L=800h^{-1}$Mpc between blue
and the orange arrows. We connect the results of $L=800$ and
$2000h^{-1}$Mpc at $k=0.1h$\,Mpc$^{-1}$ for all the redshifts. 
This is because at the baryon acoustic oscillation (BAO) scale 
($0.1h$\,Mpc$^{-1} \leq k \leq 0.3h$\,Mpc$^{-1}$), the simulation
results with $L=800h^{-1}$Mpc show a better agreement with those
obtained from the improved perturbation theory by \cite{th08}. The
initial redshift of $z_{\rm init}=99$ is high for 
$L=2000h^{-1}$Mpc with $1024^3$ particles, and hence $P(k)$ from
$L=2000h^{-1}$Mpc at low redshifts is slightly smaller than the
$P(k)$ from $L=800h^{-1}$Mpc.
Next, in connecting the $P(k)$ 
from $L=800h^{-1}$ to $320h^{-1}$Mpc, we use $P(k)$ from 
$L=800h^{-1}$Mpc up to the wavenumber where $P(k)$ agrees with 
the previous high-resolution simulation results (gray symbols) at
$z=0.35$, $1$, $3$. In this way, the connecting scales are determined
to  $k=1$, $0.6$, and $0.3h$\,Mpc$^{-1}$  for $z=0.35$, $1$, and $3$,
respectively. For the redshifts $z=0$, $0.7$, $1.5$, and $2.2$, we
interpolate the scales derived above, and for the redshifts $z=5$,
$7$, and $10$ we simply adopt the same result as that at $z=3$. The
connecting scales at each redshift are summarized in
Table~\ref{table4}. While these connecting scales are derived from the
WMAP5 model, we also use the same connecting scales in
Table~\ref{table4} for the other cosmological models. We confirmed
that the power spectra from different simulations connect smoothly
at these scales in all the
cosmological models studied in this paper.  

Figure~\ref{pk_coyote_m00} shows the power spectrum $P(k)$ in our
simulations of $L=1000h {\rm Mpc}^{-1}$ divided by that of the
 Cosmic emulator for the ten Coyote
models at $z=0$, $1$. The colored symbols correspond to the ten Coyote
models. Red symbols with error bars are for the fiducial model m00.
Here the error bars show the Gaussian errors because we have only one
realization for each Coyote cosmological model. As clearly seen in the
Figure, our simulation results agree with the Cosmic emulator within
$3 \%$ for $k=0.1-2h$\,Mpc$^{-1}$.

\section{Halofit Model}\label{sec:halo}

In the halofit model, the power spectrum consists of two terms (S03):
\beq
  \Delta^2(k) = \Delta_{\rm Q}^2(k) + \Delta_{\rm H}^2(k),
\label{pk_hf}
\eeq
where $\Delta^2(k)=k^3 P(k)/(2 \pi^2)$ is the dimensionless power
spectrum. The first term is called the two-halo term that dominates at 
large scales, whereas the second term is referred to as the one-halo
term that is important at small scales. We adopt almost the same
functional form as in S03 for both the two terms in
Equation~(\ref{pk_hf}). We use our high-resolution simulation results
to re-calibrate the model parameters of the halofit formula so as to
minimize the discrepancies.  To do so, we employ the standard
chi-squared method to find the best-fit solution:  
\beq
  \chi^2 = \sum_i \sum_{k=k_{\rm min}}^{k_{\rm max}} \sum_{z=0}^{10} W(k,z)
 \frac{\left[ P_{i, {\rm model}}(k,z)
 - P_{i, {\rm sim}}(k,z) \right]^2}  {\sigma_{i}^2(k,z)},
\label{chi-sq}
\eeq
where $P_{i, {\rm model}}$ is the model prediction, $P_{i, {\rm sim}}$
 is the simulation results, and 
 $i$ runs over the WMAP cosmological models
 shown in Table~\ref{table1}.
In Eq.(\ref{chi-sq}), it is better to include the correlation between
 the different $k$ bins at small scales for more detailed
 analysis~\citep[e.g.,][]{sco99,taka11a}.
However it is expensive to evaluate the covariance matrix of $P(k)$,
 and hence we ignore it in this paper. 
We note that we use only the six
 WMAP models in this chi-squared analysis. The remaining ten Coyote
 models are used to check the accuracy of our fitting formula. 
 We simply set the variance $\sigma_i^2=P_{i, {\rm sim}}^2$,
 and the weight function is set as follows: 
\beqa
W(k,z) &=& 10,\quad k<0.3h\,{\rm Mpc}^{-1} ~\&~ 0 \leq z \leq 3,\nonumber\\
 &=& 1, \quad 0.3h\,{\rm Mpc}^{-1} \leq k<10h\,{\rm Mpc}^{-1} ~\&~ 0 \leq z \leq 3,\nonumber\\
 &=& 0.2, \quad k<10h\,{\rm Mpc}^{-1} ~\&~ 3 < z \leq 10, \nonumber\\
 &=& 0.1, \quad k \geq 10h\,{\rm Mpc}^{-1}. \nonumber
\eeqa
The weight factor is chosen so that the
final fitting formula gives a 
better accuracy at the BAO scales at low redshifts. 

In (quasi-)linear regime, the error of $P(k)$ is given by the Gaussian
error which is the inverse of the square root of the number of
modes~\citep[e.g.,][]{fkp94}. We consider only the wavenumber
bins where the Gaussian error of $P(k)$ is less than $3\%$ for the
fitting, which correspond to $k_{\rm min}=0.02h$\,Mpc$^{-1}$.
In nonlinear regime, the non-Gaussian error arises due to the mode
coupling, but it is smaller than $5\%$ \citep[see e.g.,][]{taka09}.
While the Nyquist wavenumber is $k=10h$\,Mpc$^{-1}$, we sum up the
wavenumber up to $k=30h$\,Mpc$^{-1}$, because the nonlinear power
spectrum is reliable down to scales corresponding to the softening
length \citep[e.g.,][]{hys02}. On the other hand, we do not use the
wavenumber where the shot noise dominates the power  spectrum.
The maximum wave number is $k_{\rm max}=30h$\,Mpc$^{-1}$ at $z=0$, 
$0.35$, $k_{\rm max}=20h$\,Mpc$^{-1}$ at $z=0.7$ and $1$,
 $k_{\rm max}=10h$\,Mpc$^{-1}$ at $z=1.5$ and $2.2$,
 and $k_{\rm max}=8$, $6$, $3$, and $2h$\,Mpc$^{-1}$ at $z=3$, $5$, 
$7$, and $10$, respectively. The minimum and maximum wavenumbers
 ($k_{\rm min}$ and $k_{\rm max}$) are listed in
 Table~\ref{table4}. For all the WMAP models, the power spectrum at
 $k=k_{\rm max}$ is $10$ $(3)$ times larger than the shot noise at $z
 \leq 3$ ($5 \leq z \leq 10$). There are $35$ free parameters in our
 revised halo-model ($30$ parameters in the original model). We
 summarize the best-fit parameters in Appendix. 

\begin{figure*}
\plotone{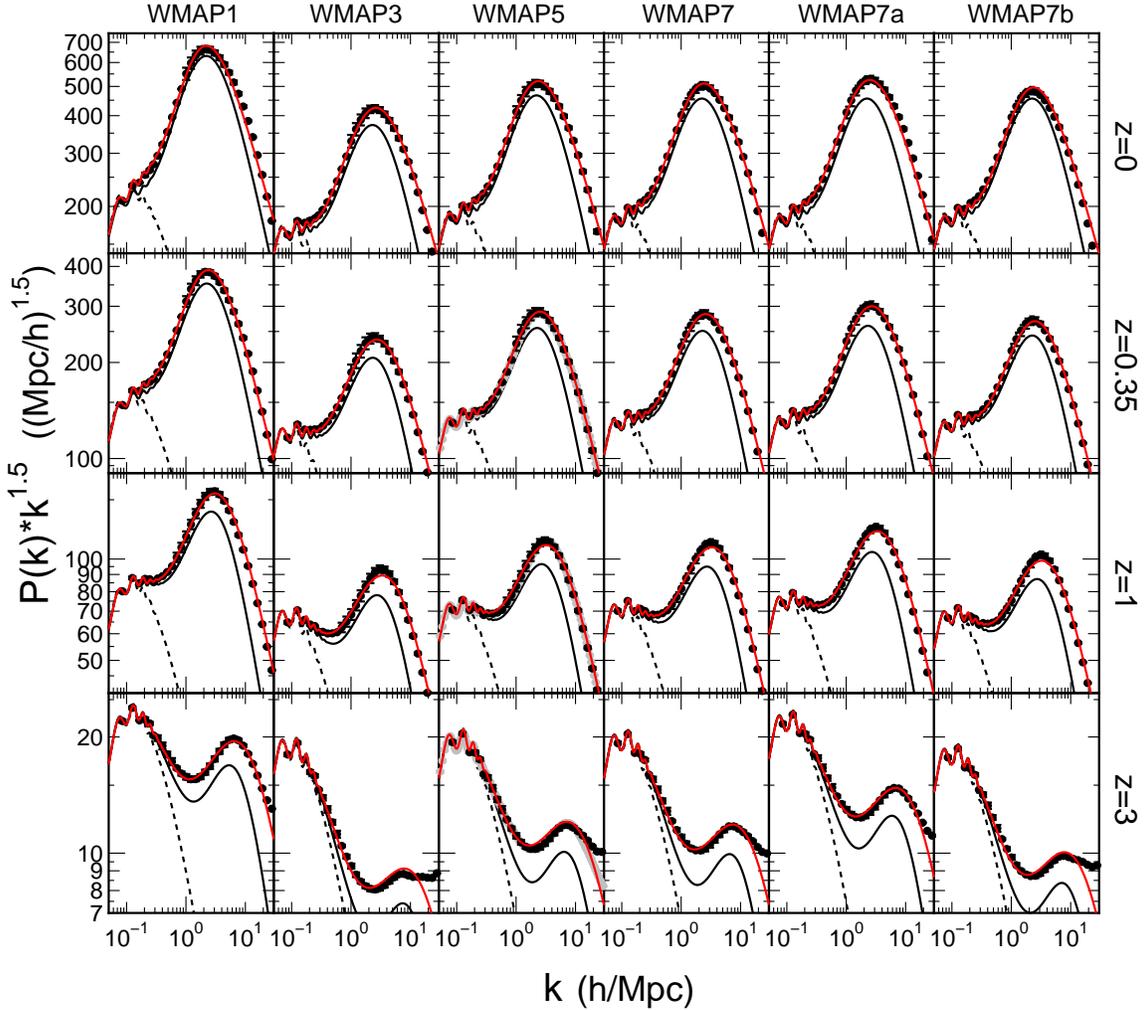}
\caption{
Power spectra $P(k)$ for the WMAP cosmological models at $z=0$, $0.35$,
$1$, and $3$. The WMAP7a,b are similar to the WMAP7 but changing the
 equation of state of dark energy ($w=-0.8,-1.2$). 
In the vertical axis, the power spectrum $P(k)$ is
multiplied by the factor $k^{1.5}$ in order to show the differences
between the simulation results and the theoretical models clearly.  
Black filled circles with the error bars plot our simulation results,
and gray symbols are the results from \cite{vn11a}.
Red solid curves show our revised halofit model (see Appendix),
whereas black solid curves show the original halofit model in
S03. Black dashed curves plot the linear power spectra. 
}
\label{pk}
\vspace*{0.5cm}
\end{figure*}

Figure~\ref{pk} shows the power spectra $P(k)$ as a function of the
wavenumber $k$ for the WMAP1, 3, 5, 7, 7a and 7b models at redshifts $z=0$,
$0.35$, $1$, and $3$. Here, to emphasize the difference between the
simulation results and the theoretical models, the power spectrum
$P(k)$ is multiplied by the factor $k^{1.5}$. Black circles with error
bars are our simulation results, whereas gray symbols in WMAP5 are the
simulation results from \cite{vn11a}. As seen in the Figure, our
simulation results agree with the results in \cite{vn11a} very
well. Red curves show our fitting function, which are significantly
better than the original halofit model in S03 shown by black
curves. As clearly seen in the Figure, the original halofit model
grossly underestimates the power spectra at smaller scales ($k \gtrsim
0.1h$\,Mpc$^{-1}$). Our model agrees with the simulation results very
well down to small scales for all the cosmological models. The
agreement of our fitting formula with simulations is better than 
$8 \%$ at $k<10h$\,Mpc$^{-1}$ for the WMAP cosmological models in the
redshift range of $0 \leq z \leq 3$. For all the WMAP cosmological
models at $k_{\rm min} \leq k \leq k_{\rm max}$, the rms deviation of
our best-fit model from the simulation results is $1.7\%$ at 
$0 \leq z \leq 10$.

\begin{figure*}
\plotone{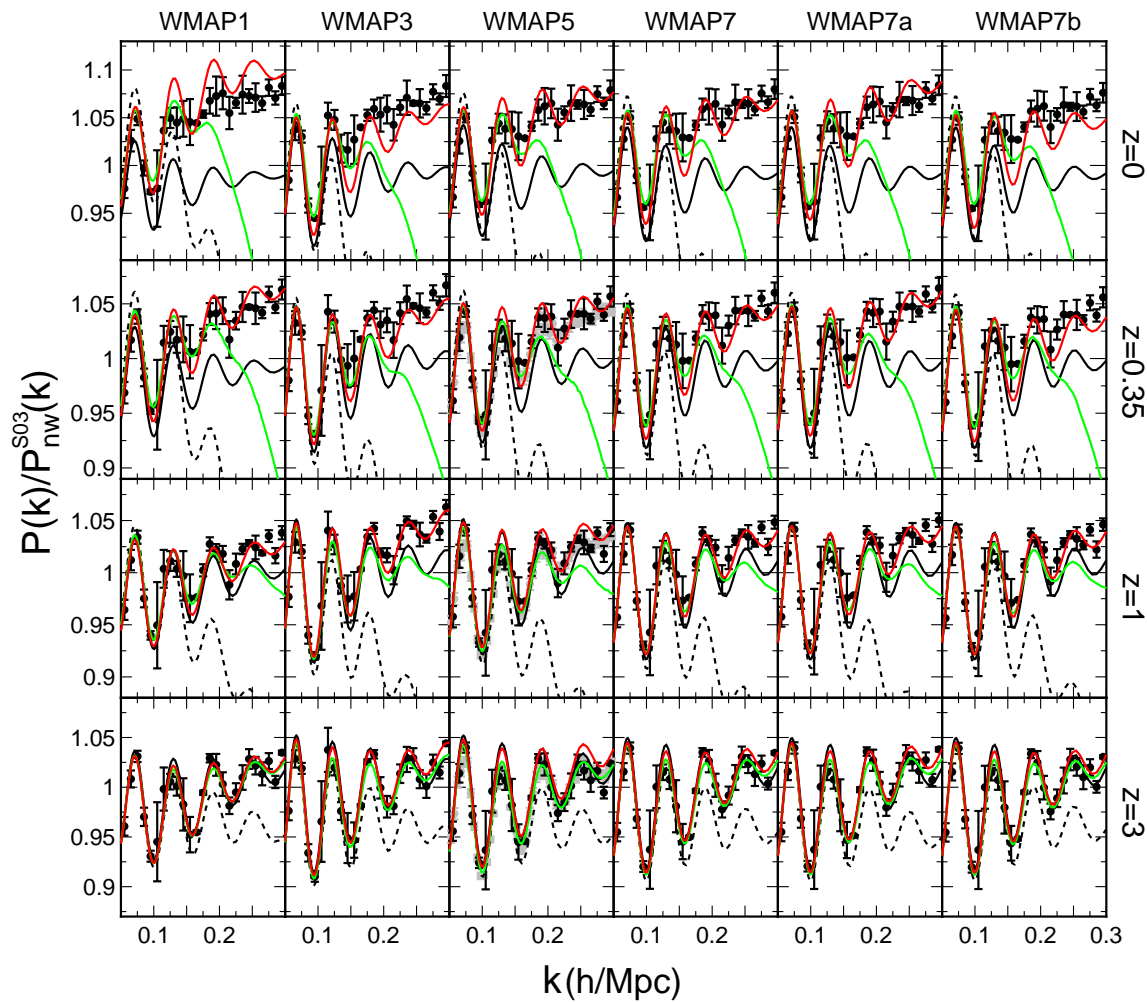}
\caption{
Similar to Figure~\ref{pk}, but focusing on the BAO scale of
$0.05h$\,Mpc$^{-1}<k<0.3h$\,Mpc$^{-1}$. The vertical axis show $P(k)$
normalized by the smooth nonlinear power spectrum 
$P_{\rm nw}^{\rm S03}(k)$ which is calculated by using a no-wiggle
fitting formula of \cite{eh98} and the original halofit model in S03
for the nonlinear correction. Again, red solid curves show our revised
halofit model, whereas black solid curves show the original halofit
model in S03. Green curves show the prediction by the
closure theory which is a higher-order perturbation theory by
\cite{th08}. Dashed curves show the linear power spectra. 
}
\label{pk_bao}
\vspace*{0.5cm}
\end{figure*}

Figure~\ref{pk_bao} shows the same results as in Figure~\ref{pk}, but
we focus on the results at the BAO scale of $0.05h$\,Mpc$^{-1}< k <
0.3h$\,Mpc$^{-1}$. In the vertical axis, the power spectrum is
normalized by the smooth nonlinear power spectrum $P_{\rm nw}^{\rm S03}(k)$, 
which is calculated by using a no-wiggle fitting formula of \cite{eh98}
with nonlinear corrections computed by the original halofit model in
S03. Green curves show theoretical predictions obtained from the
improved perturbation theory called closure theory, which efficiently
resumms a class of infinite series of higher-order perturbative
corrections~\citep{th08,ht09,taru09}. These predictions include the
corrections at the 2-loop order based on the Born approximation. The
Figure indicates that our model agrees with the simulation results better
than S03 especially at low redshifts. However, the closure theory
shows even better agreements in the quasi-linear regime. In the BAO
scales, our fitting formula reproduces simulation results within 
$4.6 \%$ for the WMAP models at $0 \leq z \leq 3$.  

\begin{figure}
\epsscale{1.1}
\plotone{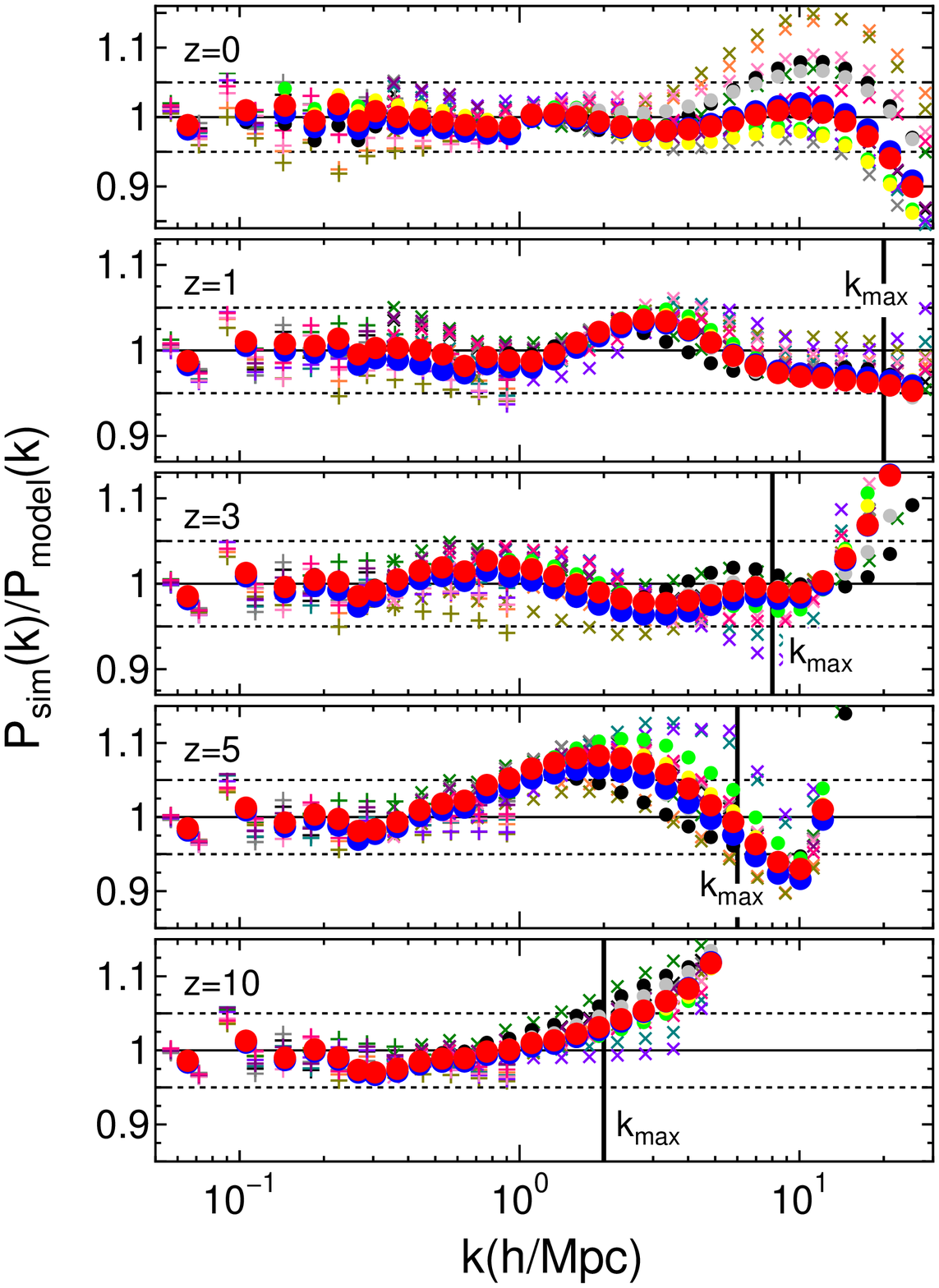}
\caption{
Our simulation results of the power spectrum $P_{\rm sim}(k)$ divided
by our improved fitting formula $P_{\rm model}(k)$ (see Appendix)
for all the cosmological models at $z=0$, $1$, $3$, $5$, and $10$  
(from top to bottom). Filled circles show results for the six WMAP
models, whereas plus and cross symbols are for the ten Coyote models. 
Blue and red big circles are WMAP5 and WMAP7. Black, green, gray and yellow
circles are results for the WMAP1, 3, 7a and 7b models,
respectively. Horizontal dotted lines indicates the relative error of
$5 \%$.  
} 
\label{our_dpk}
\vspace*{0.5cm}
\end{figure}

Figure~\ref{our_dpk} shows ratios of the measured power spectra in
our simulations to the revised fitting formula, $P_{\rm sim}(k)/P_{\rm
  model}(k)$, for all the cosmological models at $z=0$, $1$, $3$, $5$,
and $10$. Filled circles shows results for the six WMAP models,
whereas plus and cross symbols are for the ten Coyote models.
The plus and cross symbols are the results from larger (smaller)
 box sizes $L=1000 (320) h^{-1}$Mpc and are shown only for
 larger (smaller) scales of $k \leq 1 (\geq 0.3) h$\,Mpc$^{-1}$.
The horizontal dotted lines show the errors
of $5 \%$. Vertical solid lines at $z=1$, $3$, $5$, and $10$ indicate
the maximum wavenumber $k_{\rm max}$ in the chi-squared calculation in 
Equation~(\ref{chi-sq}). Note that $k_{\rm max}=30h$\,Mpc$^{-1}$ at
$z=0$. As seen in the Figure, relative errors are typically less than
$5(10)\%$ at $z\leq3(10)$ for all the cosmological models. Although we
did not include the Coyote models in our fitting, our fitting formula
reproduces simulation results for the Coyote models very well, with
the errors less than $8 \%$ for  $k \leq 1h$\,Mpc$^{-1}$ at
$z=0-10$, and $13 \%$ for $k \leq 10h$\,Mpc$^{-1}$ at $z=0-3$.
Our fiducial models (WMAP5 and 7) show better agreement,
while the other models involves slightly larger errors. At $z=0$, the
cosmological models with dark energy ($w \neq -1$) and with high
$\sigma_8$ (WMAP1) show larger errors. The cosmological models with
large (small) equation of state $w$ show the larger (smaller)
simulation results $P_{\rm sim}(k)$ than our fitting model. 
For example, for the Coyote m01 ($w=-0.816$) and  m02 ($w=-0.758$) models,
 shown as the brown and orange crosses,
 the simulation results are over $10 \%$ larger than our best fitting model
 for $k \simeq 10h {\rm Mpc}^{-1}$ at $z=0$.
At higher
redshifts $z \geq 3$, the increase of the ratio at high $k$ are due to
the shot noise. At high $z$, the errors depend mainly on the spectral
index $n_s$ since cosmological models converge to the Einstein
de-Sitter model. The models with the steep (shallow) spectral index
shows the large (small) ratio at small scales $k>1h$\,Mpc$^{-1}$.

Peacock also provided an improved halofit model which gives simply
 factor two times larger power than the original model
 for small scales $k>10h {\rm Mpc}^{-1}$
 (see his homepage\footnote{http://www.roe.ac.uk/$\tilde{~}$jap/haloes/}).
But, his model predicts a smaler power than our simulation results
 at quasi-linear scale ($k=0.1-1h\,{\rm Mpc}^{-1}$).
As clearly seen in the figure \ref{pk_wmap5}, the ratio of the simulation
 results to the halofit in S03 is not two for $k>10h {\rm Mpc}^{-1}$.
Rather, the ratio is functions of the redshift, the wavenumber and the
 cosmological parameters.

\section{Implications for Weak Lensing Predictions}\label{sec:wl}

\begin{figure*}
\epsscale{1.}
\plotone{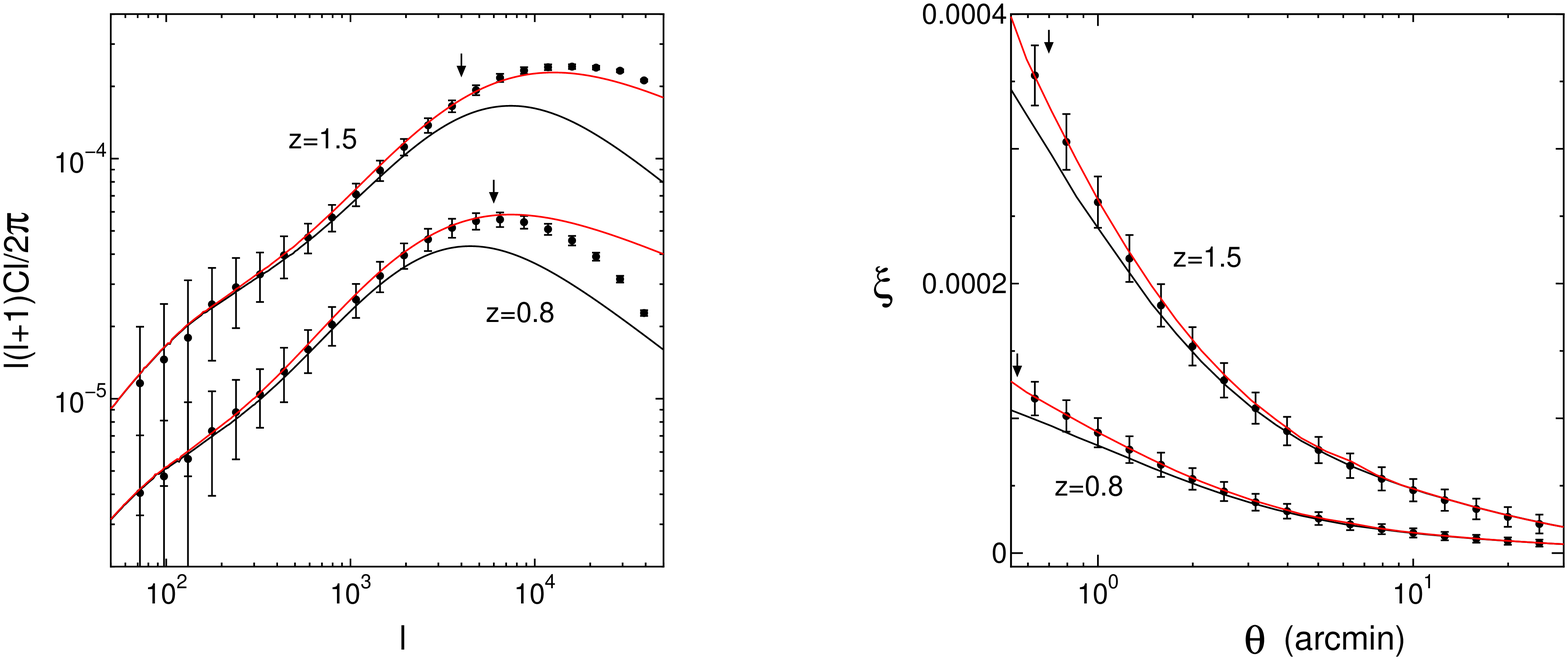}
\caption{
 Convergence power spectra ({\it left}) and correlation functions
 ({\it right}) at the source redshifts of $z_s=0.8$ and $1.5$.
 Red curves show theoretical predictions using our revised halofit
 model, and black curves show those from the original halofit model
 in S03. Filled circles with error bars plot ray-tracing simulation
 results from \cite{sato09,sato11}. Vertical arrows show the scale
 down to which the ray-tracing simulation results are accurate 
 within $5\%$. Error bars are for a $5 \times 5$~${\rm deg}^2$ survey,
 and scale as the inverse square root of the field-of-view. 
}
\label{cosmic_shear}
\vspace*{0.5cm}
\end{figure*}

\begin{figure}
\epsscale{1.}
\plotone{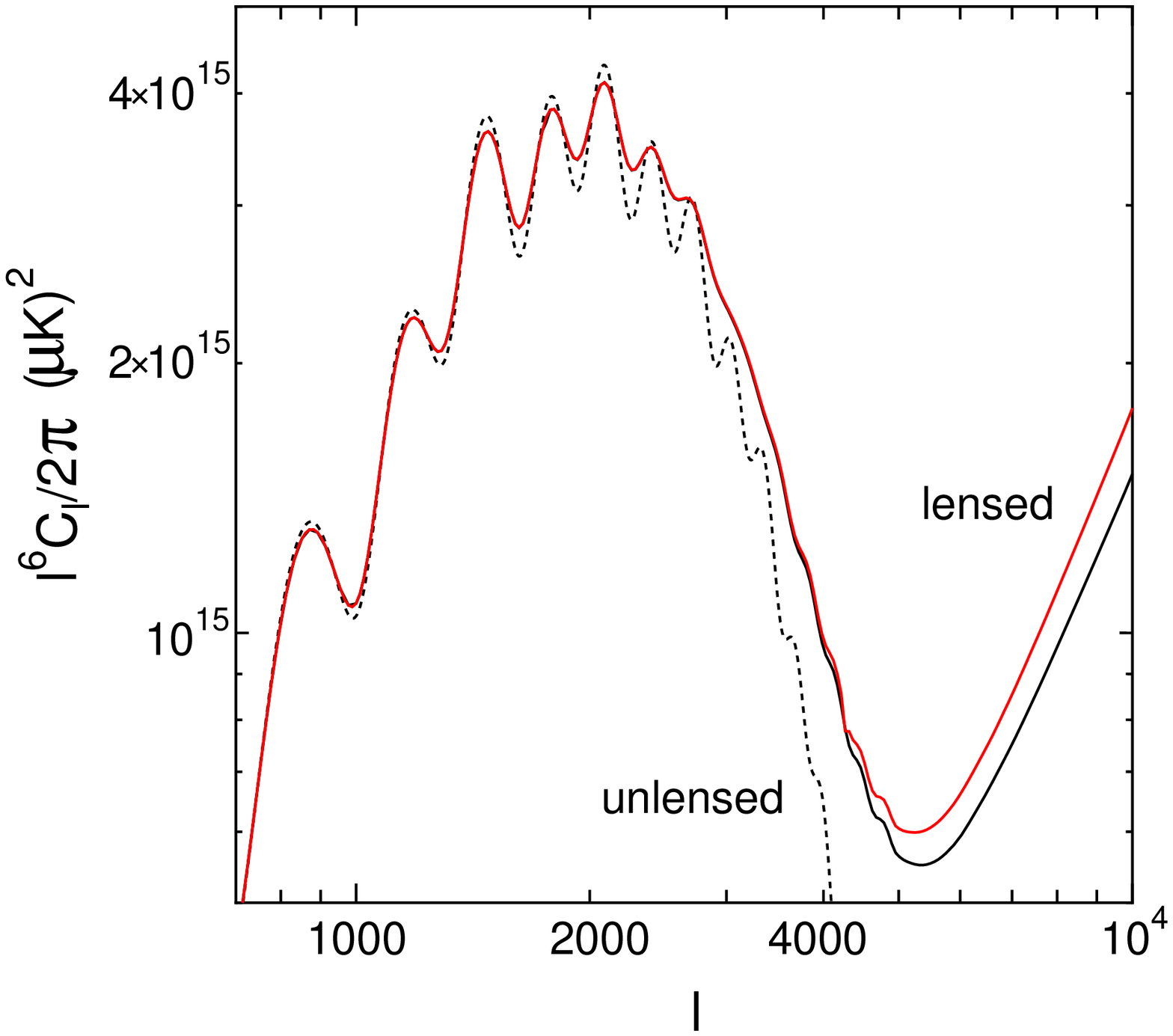}
\caption{
The lensed CMB temperature power spectrum. The red curve shows 
the prediction of the our revised halofit model, while the black
curve is that of the original halofit model in S03. The dashed curve
shows the unlensed temperature power spectrum. Note that we plot the
temperature power spectrum multiplied by $\ell^6$, i.e., 
$\ell^6 C_\ell/(2 \pi)$, not the usual convention 
of $\ell(\ell+1) C_\ell/(2 \pi)$.  
}
\label{cmb_lens}
\vspace*{0.5cm}
\end{figure}

In this section, we study how the revised model of the matter
nonlinear power spectrum affects weak lensing observables. 
Specifically, we calculate the
convergence power spectra, correlation
functions, and the CMB lensing using the new fitting formula that is
summarized in Appendix. We compare results based on our fitting
formula with those from the original halofit model in S03 as well as 
the direct ray-tracing simulation results.

Images of distant galaxies are distorted by gravitational lensing due
to intervening matter fluctuations~\citep[e.g.,][for a review]{bs01,mun08}.
The image deformation is characterized by the lensing convergence
$\kappa$ and shear $\gamma$. The convergence field is expressed as the
integration of a weighted three-dimensional density fluctuations along
the line-of-sight. Hence, the convergence power spectrum $C_\ell$ can
be expressed as a projection of the matter power spectrum weighted
with the radial lensing kernel along the line-of-sight:
\begin{equation}
 C_\ell=\int_0^{z_s} dz\,W(z,z_s)P\left(k=\frac{\ell}{\chi(z)};z\right),
\end{equation}
where $W(z,z_s)$ is a weight function.

Figure~\ref{cosmic_shear} shows the convergence power spectra and
correlation functions in the WMAP3 model at source redshifts $z_s=0.8$
and $1.5$. In both panels, red solid curves show the prediction using
our revised model of the power spectrum, and black solid curves show that
from the original halofit model in S03. Filled circles with error bars
plot direct ray-tracing simulation results obtained in  
\cite{sato09,sato11}. They used a standard ray-tracing method using
code ``RAYTRIX''~\citep{hm01}. With the $256^3$ particles in the
rectangular box of $240$ and $480h^{-1}$Mpc on each side, they
prepared $1000$ convergence and shear maps in the field of view 
$5 \times 5$ ${\rm deg}^2$. The mean and error shown in
Figure~\ref{cosmic_shear} are estimated from the $1000$
realizations. Note that the size of error bars is inversely
proportional to the square root of the sky coverage. For example, 
the Subaru HSC wide survey will observe $\sim 1500$~deg$^2$, which
suggests that expected error bars are $\sim 8$ times smaller than
those plotted here. Vertical arrows indicate the 
multipole below which their simulations are consistent with higher
resolution simulations ($512^3$ particles) within $5 \%$.
The Figure indicates that our model predictions agree with the
simulation results much better than those of the original halofit
model. This suggests that the use of the improved fitting formula
as presented in this paper is essential to extract cosmological
information from future high-resolution weak lensing measurements.

We also consider the weak lensing effect on the CMB temperature
anisotropy. Gravitational lensing by foreground matter distributions
is known to affect the light path of CMB photons coming from the last
scattering surface \citep[e.g.,][]{lc06}. As a result, the spatial
pattern of CMB temperature fluctuations is distorted, which leads to
the modification of the temperature power spectrum.  Since the lensing
deflection angle is typically a few arcmin, the temperature power
spectrum shape is significantly modified at smaller scales of
$\ell\gtrsim 1000$. Figure~\ref{cmb_lens} shows the predicted lensed CMB
temperature power spectrum. We use the CAMB code to compute the
unlensed (primordial) CMB power spectrum, which is plotted by the
dashed curve. The red solid curve is the lensed power spectrum using
our revised fitting formula, while the black solid curve is the one derived
from the original halofit model in S03. 
The fitting formula of the matter power spectrum is used 
to calculate the power
 spectrum of the deflection angle by the foreground matter distribution.
We find that our model 
enhances the power at small scales $l>4000$ by $\sim 10\%$. 

\section{Conclusion and Discussion}\label{sec:conc}

The halofit model presented in S03 has widely been used as a standard   
cosmological tool to predict the nonlinear matter power spectrum. 
However, it has been argued that the halofit model fails to reproduce
recent high-resolution simulation results such that it underestimates
the power spectrum by a few ten percent at small scales 
($k \gtrsim 0.1h$\,Mpc$^{-1}$). The difference is crucial for analysis 
of upcoming weak lensing surveys such as Subaru HSC survey, DES, and
LSST. In this paper, we have revisited the halofit model using the high
 resolution simulations for the $16$ cosmological models around the
 WMAP best-fit cosmological parameters, including the variation in the
 equation of state of dark energy.  The revised fitting formula can
 reproduce the simulation results very well in the range of
 $k<30h$\,Mpc$^{-1}$ and $0 \leq z \leq 10$. Our new fitting formula is
 summarized in Appendix, which can easily be updated from the original
 halofit model by simply replacing parameters in original model with
 new values as well as adding a few terms.    
Our revised halofit is now implemented in current version
 of CAMB\footnote{CAMB home page: http://camb.info/} (Oct. 2012),
 and hence one can easily calculate the nonlinear power spectrum $P(k)$
 and the resulting weak lensing power spectra $C_\ell$ and lensed CMB
 power spectrum $C_\ell$ in our revised halofit model using CAMB.

 We comment on effects of baryon cooling and massive neutrinos,
 both of which affect $P(k)$ at small scales. The baryon cooling
 would enhance the power at small scales by some ten percent at 
$k=10h$\,Mpc$^{-1}$ and the enhancement becomes more significant for
 smaller scales~\citep[e.g.,][]{jing06,rudd08,cas11,vd11,cas12}. However,
 the reliability of simulation results strongly relies on galaxy
 formation models they adopted. For example, \cite{vd11} showed that
 the AGN feedback can decrease the power spectrum by $>10\%$ at
 $k>1h$\,Mpc$^{-1}$. The massive neutrinos also suppress the growth
 of density fluctuation below the so-called freestreaming
 scales~\citep[e.g.,][]{bh09,bir12}. The power spectrum is suppressed
 by a few ten percent at small scales of $k \simeq 0.1h$\,Mpc$^{-1}$,
 depending on the total mass of neutrinos. Even though 
 these effects can modify the small-scale nonlinear matter power
 spectrum, an accurate knowledge of the original (dark matter only)
 nonlinear power spectra as presented in this paper is still important
 as an ingredient for building models of more realistic nonlinear power
 spectra which take these effects into account. 

Finally, while we have improved the fitting formula using the
simulations, there are also several attempts to improve the halo model
analytically. For instance, combining the perturbation theory at large
scales with the halo model at small scales,
\cite{vn11a,vn11b} and \cite{vsn12a,vsn12b} presented an improved halo
model.  On smaller scales, \cite{gio10} provided a prediction of the
power spectrum using the halo model including the effect of
substructure in the individual halo. These models also reproduce the
simulation results well and are consistent with our fitting formula.

\acknowledgments

We thank Kaiki Inoue and members in C laboratory of Nagoya University
 for useful discussion and suggestion.
M.S. and T.N. are supported by Grants-in-Aid for 
Japan Society for the Promotion of
Science (JSPS) Fellows, and A.T. acknowledges a support from 
Grant-in-Aid for Scientific Research from JSPS (No.24540257). 
This work was supported in part by Hirosaki University Grant for
 Exploratory Research by Young Scientists, by the Grant-in-Aid for Scientific
 Research on Priority Areas No. 467 ``Probing the Dark Energy through an
 Extremely Wide and Deep Survey with Subaru Telescope'',
 by the Grand-in-Aid for the Global COE Program
 ``Quest for Fundamental Principles in the Universe: from Particles
 to the Solar System and the Cosmos'' from the Ministry of Education,
 Culture, Sports, Science and Technology (MEXT) of Japan,
 by the MEXT Grant-in-Aid for Scientific Research on Innovative Areas
 (No. 21111006),
by the FIRST program "Subaru Measurements of Images and Redshifts
(SuMIRe)", World Premier International Research Center Initiative (WPI
Initiative) from MEXT of Japan, and by Grant-in-Aid for Scientific Research
from the JSPS (23740161).
Numerical computations were carried out on COSMOS provided by
 Kobayashi-Maskawa Institute for the Origin of Particles and the
 Universe, Nagoya University, SR16000 at YITP in Kyoto
 University and Cray XT4 at Center for Computational Astrophysics,
 CfCA, of National Astronomical Observatory of Japan.

\appendix

\section{Functional Form of the Revised Halofit Model}

In this Appendix, we provide the functional form of the revised halofit
model. The nonlinear power spectrum, $\Delta^2(k)=k^3 P(k)/(2 \pi^2)$,
 consist of one- and two-halo terms: 
\beq
\Delta^2(k) = \Delta_{\rm Q}^2(k) + \Delta_{\rm H}^2(k). 
\eeq
The two-halo term $\Delta_{\rm Q}^2(k)$ is given by,
\beq
\Delta_{\rm Q}^2(k) = \Delta_{\rm L}^2(k) \left[\frac{\left\{ 1+
 \Delta_{\rm L}^2(k) \right\}^{\beta_{\rm n}}}{1+\alpha_{\rm n}
 \Delta_{\rm L}^2(k)} \right] e^{-f(y)},
\label{2ht}
\eeq
where $\Delta_{\rm L}^2=k^3 P_{\rm L}(k)/(2 \pi^2)$, $f(y)=y/4+y^2/8$,
 and $P_{\rm L}(k)$ is the linear power spectrum.
The one-halo term $\Delta_{\rm H}^2(k)$ is written as 
\beq
 \Delta_{\rm H}^2(k) = \frac{\Delta_{\rm H}^{\prime 2}(k)}{1 + \mu_{\rm n}
 y^{-1} +\nu_{\rm n} y^{-2} } ~~{\rm with} 
 ~~\Delta_{\rm H}^{\prime 2}(k) = \frac{a_{\rm n} y^{3 f_1(\Omega_{\rm m})}}
 {1+ b_{\rm n} y^{f_2(\Omega_{\rm m})}+ \left[ c_{\rm n} f_3(\Omega_{\rm m})
 y \right]^{3-\gamma_{\rm n}}},
\label{1ht}
\eeq
where $y$ is the dimensionless wavenumber, $y=k/k_\sigma$.
The nonlinear scale $k_\sigma^{-1}$ is defined by
\beq
 \sigma^2(k_\sigma^{-1}) = 1 ~~{\rm with}~~
 \sigma^2(R) = \int d \ln k ~\Delta_{\rm L}^2(k) e^{- k^2 R^2}.
\eeq
The effective spectral index $n_{\rm eff}$ and the curvature $C$ are
 defined as
\beq
  n_{\rm eff} + 3 = - \left. \frac{d \ln \sigma^2 (R)}{d \ln R}
        \right|_{\sigma=1},
 ~~C = - \left. \frac{d^2 \ln \sigma^2 (R)}{d \ln R^2} \right|_{\sigma=1},
\eeq
The parameters $a_{\rm n}$, $b_{\rm n}$, $c_{\rm n}$, $\gamma_{\rm
  n}$, $\alpha_{\rm n}$, $\beta_{\rm n}$, $\mu_{\rm n}$, and 
$\nu_{\rm n}$ in Equations~(\ref{2ht}) and (\ref{1ht}) are given by 
 polynomials as a functions of $n_{\rm eff}$ and $C$.
 We determine the coefficients in the polynomials by fitting the model 
 to our simulation results, as described in Section~\ref{sec:halo}.
 The best-fit parameters are
\beqa
  && \log_{10} a_{\rm n} = 1.5222 +2.8553 n_{\rm eff} +2.3706 n_{\rm eff}^2
   +0.9903 n_{\rm eff}^3 +0.2250 n_{\rm eff}^4 - 0.6038 C   \nonumber \\
  && \hspace{5cm} +0.1749 \Omega_{\rm w}(z) \left( 1+w \right),
  \label{bf-params1} \\
  && \log_{10} b_{\rm n} = -0.5642 +0.5864 n_{\rm eff} +0.5716 n_{\rm eff}^2
    -1.5474 C+0.2279 \Omega_{\rm w}(z) \left( 1+w \right),
  \label{bf-params2}  \\
  && \log_{10} c_{\rm n} = 0.3698 +2.0404 n_{\rm eff} +0.8161 n_{\rm eff}^2
    +0.5869 C,  \\
  && \gamma_{\rm n} = 0.1971-0.0843 n_{\rm eff} +0.8460 C, \\
  && \alpha_{\rm n} = \left| 6.0835 +1.3373 n_{\rm eff}-0.1959 n_{\rm eff}^2
     -5.5274 C \right|, \label{bf-params3}  \\ 
  && \beta_{\rm n} = 2.0379 -0.7354 n_{\rm eff}+0.3157 n_{\rm eff}^2
     +1.2490 n_{\rm eff}^3 +0.3980 n_{\rm eff}^4 -0.1682 C, \\
  && \mu_{\rm n} = 0, \label{bf_params5}  \\
  && \log_{10} \nu_{\rm n} = 5.2105+3.6902 n_{\rm eff},
\label{bf-params4}
\eeqa
where $\Omega_{\rm w}(z)$ is the dark energy density parameter at
redshift $z$. The last terms in Equations~(\ref{bf-params1}) and
(\ref{bf-params2}) represent small correction terms for dark energy $w
\neq -1$. We use the absolute value of $\alpha_{\rm n}$ in
Equation~(\ref{bf-params3}) to avoid divergence in the two-halo term
(Equation~(\ref{2ht})). Finally, $f_{1,2,3}(\Omega)$ in
Equation~(\ref{1ht}) are the same as in S03
\beq
 f_1(\Omega_{\rm m}) = \Omega_{\rm m}^{-0.0307}, ~f_2(\Omega_{\rm m}) =
 \Omega_{\rm m}^{-0.0585}, ~f_3(\Omega_{\rm m}) = \Omega_{\rm m}^{0.0743},
\eeq
where $\Omega_{\rm m }$ is the matter density parameter at redshift $z$. 

In summary, one can easily revise the original halofit model in S03 by
replacing the parameters
 $a_{\rm n}$, $b_{\rm n}$, $c_{\rm n}$, $\gamma_{\rm n}$, $\alpha_{\rm n}$,
 $\beta_{\rm n}$, $\mu_{\rm n}$, and $\nu_{\rm n}$ to those listed in
 Equations~(\ref{bf-params1})-(\ref{bf-params4}). 


\clearpage

\end{document}